\documentclass[5p,a4paper,11pt,lmodern,jnm]{elsarticle}
\usepackage{graphicx,xcolor,caption,subcaption,float,amsmath,amssymb,amsfonts,amsthm,amsbsy,amstext,stmaryrd}
\usepackage{siunitx}
\usepackage{xfrac}
\usepackage{hyperref}
\hypersetup{colorlinks=true}
\usepackage[english]{babel}
\usepackage{balance}

\biboptions{square,comma,super,sort&compress}

\begin{document}

\begin{frontmatter}
\title{On the Ground State of the U-Mo System}

\author[cab1]{E.L.~Losada$^\ast$}
\author[cab2]{J.E.~Garc\'es}  
\address[cab1]{GCCN, Centro At\'omico Bariloche, Comisi\'on Nacional de Energ\'ia At\'omica, Argentina}
\address[cab2]{GIA, Centro At\'omico Bariloche, Comisi\'on Nacional de Energ\'ia At\'omica, Argentina}

\begin{abstract}
The ground state of the U-Mo system is studied in this work using an evolutionary algorithm coupled with \textit{ab initio} calculations. This methodology is applied to several compositions, with a more detailed examination at the U-rich side. A new ground state is identified within an approach that considers structures for up to 3 formula units per cell. A supercell scheme based on the pure phase structure of each element is used for dilute compositions. The ground state at a pressure of 1 atm. is characterized by only two stable compounds: $\Omega$-phase and a new phase, named $\alpha$-U$_{15}$Mo in this work, with a structure determined by adding a substitutional Mo impurity to a $1\times 2\times 2$ supercell of $\alpha$-U. A range of immiscibility of U in bcc-Mo based structures is extended from 0 to nearby 50 at. \% of U. The new $\alpha$-U$_{15}$Mo phase comprising the ground state has similar features to those of the $\alpha'$-phase formed in quenched samples at room temperature. The internal relaxation of U atoms in this phase resembles the movements due to charge density waves observed in orthorhombic $\alpha_1$-U. The ground state of the U$_{15}$Mo compound at T=0 K can not be fully identified due to the charge density wave associated with $\alpha$-U. That structure could be a supercell based on the structurally undefined $\alpha_3$-U phase.
\end{abstract}

\end{frontmatter}

\long\def\symbolfootnote[#1]#2{\begingroup%
\def\thefootnote{\fnsymbol{footnote}}\footnote[#1]{#2}\endgroup}
\long\def\symbolfootnotemark[#1]{\begingroup%
\def\thefootnote{\fnsymbol{footnote}}\footnotemark[#1]\endgroup}
\long\def\symbolfootnotetext[#1]#2{\begingroup%
\def\thefootnote{\fnsymbol{footnote}}\footnotetext[#1]{#2}\endgroup}
\symbolfootnotetext[1]{Corresponding author email: losada@cab.cnea.gov.ar}

	
\section{INTRODUCTION}\label{sec.introd}
(U,Mo) and (U,Zr) solid solutions have technological relevance due to its potential use as nuclear fuel in advanced GenIV reactors \citep{Kim}. (U,Mo) is also a promising candidate for the development of high-density Uranium fuels to be used in Material Testing Reactor (MTR) \citep{Snelgrove}. Both metallic fuels have a number of advantages over oxide fuels in relation to their better thermal conductivity and evolution at high burn-up. 

Both above mentioned systems are known to have only one stable compound at room temperature, which are the U$_2$Mo in the $I4/mmm$ space group \citep{LANDA201331} and the $\delta$-UZr$_2$ phase with $P6/mmm$ space group \citep{Landa2011132,XIONG,MCKEOWN}. These phases gradually transforms into a bcc solid solution with increasing temperature \citep{Massalski}. 

The low-temperature phase of pure uranium metal is called  alpha-Uranium ($\alpha$-U) which is stable below 935 K. It still remains as an intriguing material as it is the only three dimensional single element, so far, which exhibits spontaneous  phase transitions at ambient pressure related to charge-density waves (CDW). They are called $\alpha_1$, $\alpha_2$ and $\alpha_3$ taking place at temperatures of 43 K, 37 K and 23 K, respectively \citep{Tindall,Fisher,Steinitz}. The CDW of $\alpha_1$–phase was explained theoretically by Fast \textit{et al.} \citep{Fast} to be a result of a Peierls-like transition involving the formation of pairs of U atoms along the b axis. Nowadays, the problem of CDW in $\alpha_1$–U phase still attracts attention. 
Qiu et al. \citep{Qiu} presented a detailed study of this phase through a first-principles total-energy minimization using the conjugate gradient algorithm. However, due to the amount of atoms involved in the CDW transitions associated to $\alpha_2$ and $\alpha_3$ phases, no attempts were found in the literature to describe theoretically these phases. The last one has a primitive cell with volume of $\sim 6000\AA^3$ \citep{Smith,Lander1994}. In addition, several physical quantities at lower temperatures show anomalies such as a controversial superconductivity transition (T$_{SC}$ = 0.78 to 0.48 K) \citep{Marmeggi,Raymond}.

Metastable phases based on the orthorhombic $\alpha$–U structure are observed in quenched samples with Mo content below 12 at.\%. They were named $\alpha'$ and $\alpha"$ by Lehmann and Hills \citep{Lehmann}. The $\alpha'$ phase is found for a Mo content below 6.25 at.\% It has an orthorhombic structure in which the \textit{b} parameter is contracted with respect to the pure  $\alpha$–U structure. The phase $\alpha"$ is found in a range of compositions between 6.25 at.\% and 11 at.\% of Mo, being a monoclinic distortion of $\alpha'$ \citep{Tangri}. 

Although the U-X (X=Mo,Zr) systems are well characterized above room temperature, no experimental or theoretical information is available regarding the behavior of stable or metastable phases at low temperatures. It was assumed in modeling efforts that the stable compounds at room temperatures are the same as those at T = 0 K, since there was no information available to contradict this hypothesis. Thereby, accurate theoretical calculations of enthalpy of forma-\linebreak tion assumed one compound with the structure I4/mmm in the ground state (GS) of U-Mo system \citep{LANDA201331}. Only one compound was also used to model the U-Mo \cite{XIONG} and U-Zr phase diagrams \citep{Zhang201015}. 

However, it was recently found that the U$_2$Mo composition with the $I4/mmm$ structure is unstable upon applying a distortion related to the $C_{66}$ elastic constant \citep{Losada2015,Losada2016}. Furthermore, the existence of a new structure with the hexagonal $P6/mmm$ space group (coincidentally resembling the structure of $\delta$-UZr$_2$ phase) was predicted by using theoretical \textit{ab initio} calculations. It was called $\Omega$-phase and it should show up in the phase diagram at least in the vicinity of absolute zero as it is $\sim 4.7$ mRy more stable than the $I4/mmm$, previously assumed as the GS for this composition \citep{Losada2015,Losada2016}. The total energy of both structures were computed for the sake of comparison in the primitive unit cell. This result was confirmed by Chen et al. \citep{KeChen}. 

The unexpected result found in the U$_2$Mo compound emphasizes the need of a careful examination of the GS in the U-Mo system. However, the accurate depiction of the GS is one of the main problems in Material Science remaining unsolved for a long time until new theoretical methods were developed. One of these methodologies is the cluster expansion of the total energy used in \textit{ab initio} thermodynamics. Another is based on the usage of evolutionary algorithms coupled with \textit{ab initio} calculations. The successful implementation of both methodologies relies on the accurate calculation of the total energy of stable and metastable structures. There will always be doubt that the obtained GS is the correct one because many structures remain closely located near the minimal energy for each composition. A couple of issues should be considered in detail related to this topic: 
1) The ability and precision of \textit{ab initio} programs to achieve the optimal structural relaxation. Sometimes, a tiny distortion of atoms positions can lead to different space group symmetries. This is what happened in the case of the U$_2$Mo $\Omega$-phase where the structure with the P6/mmm space group turned out to be only 0.1 mRy more stable than the Pmmn structure in the U$_2$Mo compound \citep{Losada2015,KeChen,Wang}. 
2) The modeling method selected in the DFT code should properly describe the physics of the system; otherwise, the structure could evolve to a situation different to the naturally occurring one.

There are three possibilities for the U-Mo GS compatible with the third law of Thermodynamics \citep{Abriata}: 
i) a complete phase separation in pure Mo and U elements, 
ii) a phase diagram remaining unchanged towards T = 0 K and 
iii) the existence of one or several ordered compounds for different compositions becoming metastable or not with increasing temperature. It is important to comment here that there is still no information available in the scientific literature between the metastable phases and the GS in the U-Mo system.

The aim of this research is to find out which of these three options occurs in the U-Mo system and to identify the structures belonging to the GS. For that purpose, the USPEX evolutionary algorithm \citep{Uspex1,Uspex2,Uspex3} coupled with the Quantum Espresso code (QE) \citep{QE2009,QE2017} is implemented in this work in order to find the potential candidate structures to constitute the GS at different compositions in the U-rich side of the phase diagram.  As a final step, the Wien2k code \citep{Wien2k} is used to compute the total energy of the previously obtained structures and to achieve an accurate comparison between them.

This work is organized as follows: Section \ref{sec.detailscalc} describes the methodology used to select the set of structures studied in this work. Section \ref{sec.structProp}  presents the proposed GS for the U-Mo system. Section \ref{sec.structGS}  discusses structural properties of the GS. Conclusions are given in Section \ref{sec.conclusions}.

\section{DETAILS OF CALCULATIONS}\label{sec.detailscalc}
USPEX is one out of many evolutionary algorithms used to predict stable and metastable structures knowing  only its chemical composi-\linebreak tion \citep{Uspex1,Uspex2,Uspex3}.

\begin{figure}[h]
	\centering
	\begin{subfigure}[b]{0.35\columnwidth}
		\includegraphics[width=\textwidth]{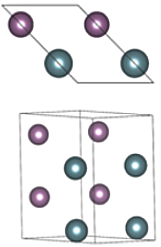}	\caption{}
		\label{fig:parecidasA}
	\end{subfigure}
	\begin{subfigure}[b]{0.3\columnwidth}
		\includegraphics[width=\textwidth]{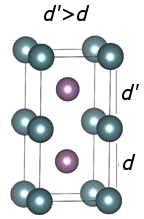}
		\caption{}
		\label{fig:parecidasB}
	\end{subfigure}
	\begin{subfigure}[b]{0.3\columnwidth}
		\includegraphics[width=0.6\textwidth]{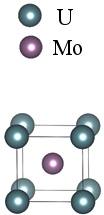}
		\caption{}
		\label{fig:parecidasC}
	\end{subfigure}
	\caption{ Structures $(a)$ and $(b)$  proposed by USPEX code evolving to the same structure $(c)$ with the CsCl prototype.}
	\label{fig:parecidas}
\end{figure}

One of the major difficulties encountered in this work with this methodology is related to the accurate determination of optimized structures. The usage of very precise convergence parameters may be time consuming and is no guarantee of success in the correct determination of the optimal structure symmetry. The case of the Pmmn structure \citep{Wang} which could not be solved into the P6/mmm structure of the U$_{2}$Mo $\Omega$-phase is a clear example of this situation. Another one found in this work is the case of a trigonal unit cell evolving to a simple cubic one, for which the variable cell relaxation never reaches angles of 90 degrees. Fig. \ref{fig:parecidas} depicts this situation where two structures belonging to different spacegroups evolve to the same structure with the B2 Strukturbericht designation, Pm-3m spacegroup and CsCl prototype. Therefore, the methodology implemented in this work was not the standard one. The USPEX coupled with QE code was used to identify the best set of potential structures followed by a visual inspection in order to avoid repetition of similar structures. The final structure is selected based on symmetry criteria and a careful visual analysis made using XcrysDen and Vesta codes. A subsequent structural optimization and relaxation of internal positions was carried out by using the Wien2k code because the energy differences between candidate structures were too small for some compositions. 

In the particular region of low solute concentration, structures were introduced by hand to simulate the presence of one or two impurities in a matrix with the structure corresponding to the pure elements phases. Therefore, at the Mo-rich side, a $3\times3\times3$ supercell with the pure Mo-bcc structure was considered as a base for the compositions UMo$_{53}$ and U$_{2}$Mo$_{52}$; and at the U-rich side, the orthorhombic $\alpha$-U phase structure was used as the building block of a $1\times2\times2$ supercell for U$_{15}$Mo and U$_{14}$Mo$_{2}$ compositions. Details of these two structures will be given below.

The Wien2k code \citep{Wien2k} uses the full-potential LAPW+lo method that makes no shape approximation to the potential or density. Electronic exchange-correlation interactions were treated\linebreak within the generalized gradient approximation of Perdew, Burke and Ernzerhof \cite{Pedrew1996}, as no experimental or theoretical evidence of strong correlations was found in the system studied here. The radii of the atomic spheres (R$_{MT}$) selected for U and Mo were R$_{MT_{U}}$ = 2.3 a.u. and R$_{MT_{Mo}}$  = 1.8 a.u., respectively. In order to describe the electronic structure of all the atoms and their orbitals, the APW + lo basis set was selected. Local orbital extensions were included to describe the semicore states by means of the method \linebreak(APW + lo), as implemented in the WIEN2k program. The cut-off parameter that controls convergence in the expansion of the solution to the Kohn–Sham equations was chosen to be Rkmax=8.
The maximum $l$ values for partial waves used inside the atomic spheres and for the non-muffin-tin matrix elements were selected to be $l_{max}$ = 10 and $l_{max}$ = 4, respectively. The charge density cut-off $G_{max}$ was selected as $22Ry^{1/2}$. The k-points at which the Brillouin zone was sampled, were determined considering its density in the reciprocal-space, as many structures of very different sizes were compared. For this reason, typical k-points spacing were chosen to be around 0.04. k-space integration was calculated using the modified tetrahedron-method \citep{Blochl}. The iteration process is repeated until the calculated total energy converges to less than $1\times10^{-6}$ Ry/cell, and the calculated total charge converges to less than $1\times10^{-6}$ e/cell. The mini LAPW script implemented in the Wien2K package was employed to calculate the internal parameters of the crystal structures. For completeness of this study,  spin-polarized calculations and spin-orbit interactions were included, but later on they were dismissed as no significant effects were observed in the relative stability of structures.

\section{RESULTS AND DISCUSSIONS}\label{sec.results}
The most relevant motivations for fully characterizing the structures 
comprising the GS of a system are related to: 
i) the identification of stable and metastable structures for all compositions finding out novel phases at low temperatures, 
ii) the improvement of phase diagram through \textit{ab initio} thermodynamics or CALPHAD methodology, 
iii) the discovery of unexpected properties like the Peierls distortion induced by deformation in the U$_2$Mo compound, the CDW in the U$_2$Ti structure \citep{Kaur}, or the superconducting phase transition of pure U \citep{Fast}, and 
iv) the understanding of thermophysical properties at T=0 K in order to be extrapolated, using \textit{ab initio} Molecular Dynamics \citep{chakraborty}, at the experimental condition of interest in a nuclear reactor.

This section presents the results of each studied composition in the U-Mo system at T=0 K.

\subsection{Structures and Phase stability of the U-Mo system}\label{sec.structProp}
\begin{figure*}[!h]
	\center\fcolorbox{black}{white}{\includegraphics[width=0.80\textwidth]{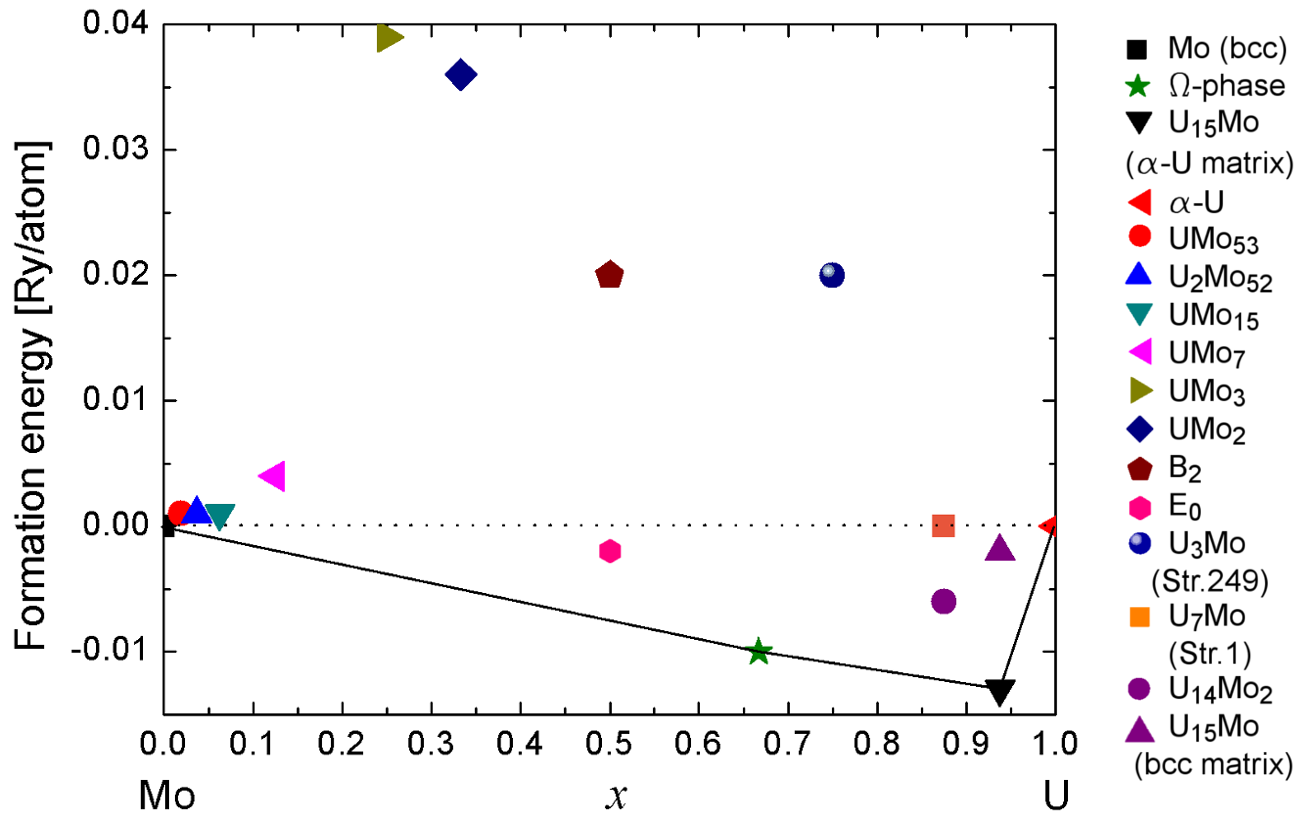}}\\
	\caption{Convex hull of the U-Mo system showing a GS formed by two stable structures: $\Omega$-phase and the U$_{15}$Mo based on $\alpha$-U phase matrix. A range of immiscibility of U in bcc-Mo based structure extends from compositions $x=0$ to approximately $x=0.5$.  bcc-Mo and orthorhombic $\alpha$–U are used as reference states. Only structures with formation energies below 0.04 Ry/atom are shown.}\label{fig:convexhull}
\end{figure*}

Global search for candidates to the GS was carried out at several stoichiometries, namely: \linebreak UMo$_{53}$, U$_{2}$Mo$_{52}$, UMo$_{15}$, UMo$_{7}$, UMo$_{3}$, UMo$_{2}$, UMo, U$_{2}$Mo, U$_{3}$Mo, U$_{7}$Mo, U$_{14}$Mo$_{2}$ and U$_{15}$Mo. The combination of USPEX plus QE code was used mainly for U compositions:
\begin{equation}\label{eq.composition}
x = \frac{n_U} {(n_U+n_{Mo})} \geq 0.5 \,,
\end{equation}
where $x$ measures the ratio of U atoms number ($n_U$) to total number of atoms ($n_U+n_{Mo}$). The implemented methodology is described in Section \ref{sec.detailscalc}. The formation energy per atom is calculated to compare the stability of all considered structures related to pure compositions. It is defined as follows:
\begin{equation}\label{eq.formE}
\Delta E_{f_{Str.}}=\mathcal{E}_{Str.}\, -x\, \mathcal{E}_{\alpha U}-(1-x)\,\mathcal{E}_{bccMo} \,,
\end{equation}
where $E_{f_{Str.}}$ is the formation energy of structure under analysis relative to pure compounds; $x$ is the U composition of structure, previously defined in Eq. \ref{eq.composition}; $\mathcal{E}_{Str.}$, $\mathcal{E}_{\alpha U}$ and $\mathcal{E}_{bccMo}$ are the total energy per atom in the unit cell of the structure and pure phases found at 0 K, in this case the orthorhombic $\alpha$-U and bcc-Mo. The relative stability of studied compounds and the resulting ground state are visualized by means of  construction of the thermodynamic convex hull, depicted in Fig. \ref{fig:convexhull}. It is a diagram with data pinpointing the relationship between formation energy and composition for each considered structure. A trace connecting the points corresponding to the minimum energy phases with the pure ones, highlights the structures belonging to the GS. Thereby, each structure over the GS is stable with respect to decomposition into neighboring structures. 


The GS of U-Mo system at a pressure of 1 atm is characterized by only two stable compounds: $\Omega$-phase and the  U$_{15}$Mo structure, as it is shown in Fig. \ref{fig:convexhull}.  A range of immiscibility of U in bcc-Mo based structures is extended from compositions $x=0$ to approximately $x=0.5$. All studied structures in the range $x < 0.5$ have positive formation energies and, consequently, are unstable. For compositions $x > 0.5$, there are structures having negative formation energies. They are stable but metastable with respect to the convex hull. The obtained GS is compatible with the experimental information from the phase diagram \cite{Massalski} and also it is in agreement with the Third Law of Thermodynamics. 

\begin{figure}[!h]
	\centering
	\begin{subfigure}[b]{0.4\columnwidth}
		\includegraphics[width=\textwidth]{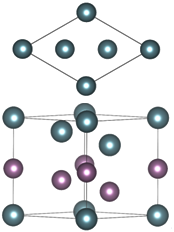}
		\caption{E0}
		\label{fig:50a}
	\end{subfigure}\hspace{1cm}
	\begin{subfigure}[b]{0.24\columnwidth}
		\includegraphics[width=\textwidth]{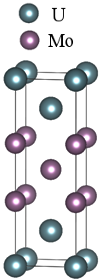}
		\caption{E3}
		\label{fig:50b}
	\end{subfigure}\\
	\begin{subfigure}[b]{0.25\columnwidth}
		\includegraphics[width=\textwidth]{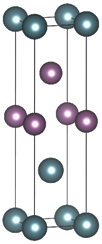}
		\caption{B11}
		\label{fig:50c}
	\end{subfigure}\hspace{1.3cm}
	\begin{subfigure}[b]{0.25\columnwidth}
		\includegraphics[width=\textwidth]{50c.png}
		\caption{B2}
		\label{fig:50d}
	\end{subfigure}
	\caption{Lower energy structures for Mo-50 at.\%U (UMo).}\label{fig:50per}
\end{figure}

\begin{figure}[h]
	\center\fcolorbox{black}{white}{\includegraphics[width=0.95\columnwidth]{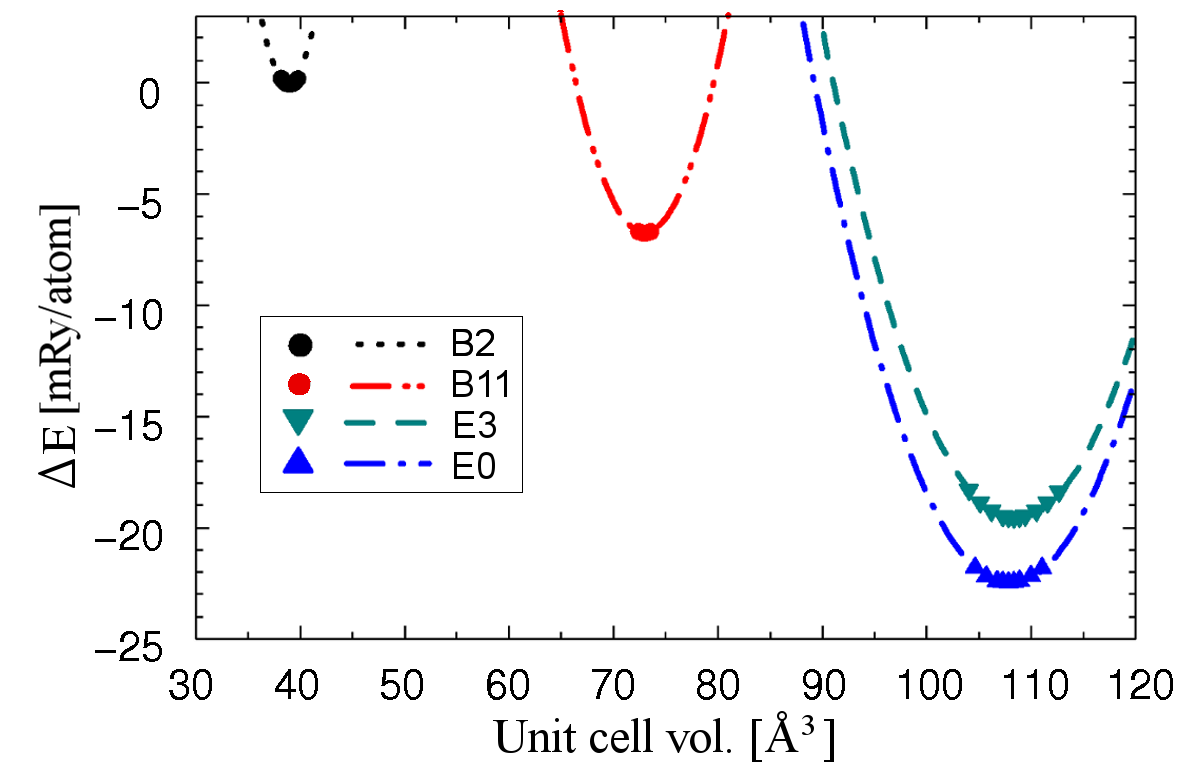}} 
	\caption{Total energy difference per atom comparison as a function of the unit cell volume variation between candidates for the UMo composition. Reference energy corresponds to the optimized B2 structure. The ground state is the structure named E0 in this work. }\label{fig:CompVol}
\end{figure}

The structures of the GS and most of the metastable compounds are a consequence of the competition between U-Mo-U blocks, as in the case of the $\Omega$-phase, and infinite chains of U and Mo, as in the case of the U$_{15}$Mo compound. Their lattice parameters are shown in Table \ref{tb:GSstructures}. Therefore the metastable structures found in this work, limited to 3 formula units (f.u.) per cell, are the result of a competition between pure planes of Mo or U atoms and chains formed by alternating U and Mo atoms, as it is observed for composition $x = 0.5$. Fig. \ref{fig:50per} shows the analyzed structures at this composition, which are: i) B2 formed by consecutive planes of U and Mo, ii) B11 composed by two alternating planes of each atom, iii) the structure called E3 in this work which has three planes of U and Mo atoms, and iv) structure called E0 composed by linear chains of alternated U and Mo atoms. E0 corresponds to the metastable GS at $x = 0.5$. It is worth noting that the B11 structure (P4/nmm space group), after internal positions relaxation, undergoes a structural modification to P4mm space group where the middle plane with Mo atoms gets moved to a final position different of the original one with $z=0.5$. The relative stability between structures is presented in Fig. \ref{fig:CompVol}.

\begin{table}[]
\centering
\begin{tabular}{|l|}
\hline 
\hspace{2.5cm}$\Omega$-U$_2$Mo \\
\hline 
Space group: \#191 P6/mmm  \\
Lattice parameters: a=b=4.8207$\AA$ \\
c=2.7642 $\AA$\\
Wyckoff possitions:  \\
Mo (1a) (0, 0, 0) \\
U (2d) (1/3,2/3,1/2) \\
\hline 
\hspace{2.8
	cm}U$_{15}$Mo \\
\hline 
Space group: \#25 Pmm2  \\
Lattice parameters: a=4.9563$\AA$\\ b=5.6427$\AA$\\
 c=11.5214 $\AA$\\
Wyckoff possitions:  \\
Mo (1a) (0, 0, 0) \\
U1 (1d) (1/2,1/2,z) z=0.394 \\
U2 (1b) (0,1/2,z) z=0.997 \\
U3 (1c) (1/2,0,z) z=0.402 \\
U4 (1d) (1/2,1/2,z) z=0.899 \\
U5 (1b) (0,1/2,z) z=0.497 \\
U6 (1c) (1/2,0,z) z=0.907 \\
U7 (1a) (0,0,z) z=0.499 \\
U8 (2h) (1/2,$\pm$ y,z) y=0.759 z=0.146 \\
U9 (2m) (0,$\pm$ y,z) y=0.761 z=0.247 \\
U10 (2h) (1/2,$\pm$ y,z) y=0.751 z=0.651 \\
U11 (2m) (0,$\pm$ y,z) y=0.746 z=0.749 \\
\hline 
\end{tabular}
\caption{Structural information of phases comprising the U-Mo system GS.}\label{tb:GSstructures}
\end{table}

 \begin{figure}[htb]
	\centering
	\begin{subfigure}[b]{0.4\columnwidth}
		\centering
		\includegraphics[width=\textwidth,keepaspectratio]{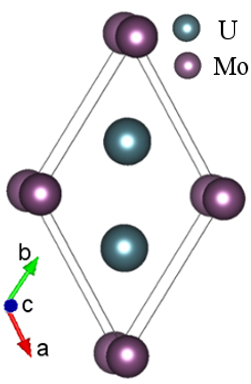}
		\caption{$\Omega$-phase (U$_2$Mo).}
		\label{fig:Omega}
	\end{subfigure}
	\begin{subfigure}[b]{0.4\columnwidth}
		\centering
		\includegraphics[width=\textwidth,keepaspectratio]{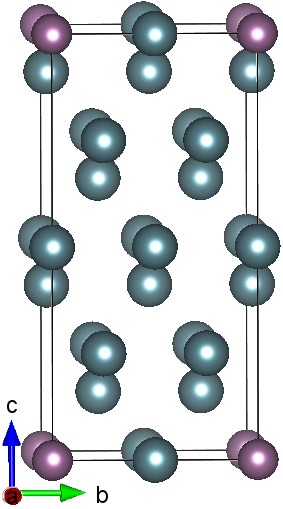}
		\caption{U$_{15}$Mo.}
		\label{fig:U15Mo}
	\end{subfigure}
	\caption{Structures constituting GS of the U-Mo system: the $\Omega$–phase (left) and U$_{15}$Mo in  a $1\times2\times2$ supercell based on $\alpha$-U phase matrix (right).}\label{fig:GSstructures}
\end{figure}


 \begin{figure}[htb]
	\centering
	\begin{subfigure}{0.4\columnwidth}
		\centering
		\includegraphics[width=0.9\textwidth,keepaspectratio]{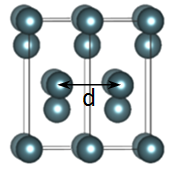}
		\caption{Twice $\alpha$-U.}
		\label{fig:Dispalpha}
	\end{subfigure}
	\begin{subfigure}{0.4\columnwidth}
		\centering
		\includegraphics[width=\textwidth,keepaspectratio]{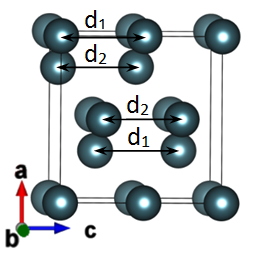}
		\caption{    $\alpha_1$-U $(d_1 > d_2)$.}	
		\label{fig:Dispalpha1}
	\end{subfigure}\\
	\begin{subfigure}{0.4\columnwidth}
		\centering
		\includegraphics[width=\textwidth,keepaspectratio]{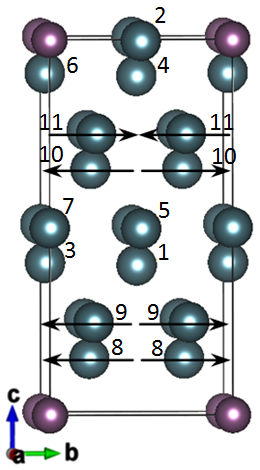}
		\caption{U$_{15}$Mo.}
		\label{fig:DispU15Mo}
	\end{subfigure}
	\begin{subfigure}{0.4\columnwidth}
		\includegraphics[width=\textwidth,keepaspectratio]{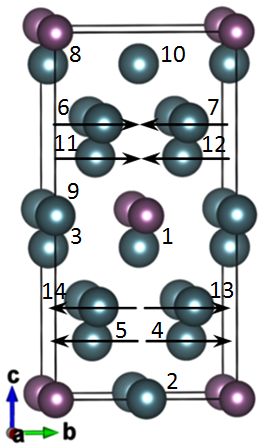}
		\caption{U$_{14}$Mo$_2$.}
		\label{fig:DispU14Mo2}
	\end{subfigure}\\
	\caption{Internal U atoms displacements observed in U$_{15}$Mo and U$_{14}$Mo$_2$ related to the presence of CDWs in $\alpha_1$-U. 
	A global translation of coordinates was made in $\alpha$-U and $\alpha_1$-U to be able to compare similar atomic displacements in different structures.}\label{fig:Displacements}
\end{figure}
\subsection{Structural features of the GS structures.}\label{sec.structGS}

The results of this work show that only two structures belongs to the U-Mo system GS: the $\Omega$-phase and U$_{15}$Mo (in $\alpha$-U phase matrix), both depicted in Fig. \ref{fig:GSstructures}. 

\subsubsection{$Omega$-phase}
The $\Omega$–phase has recently been studied in detail \cite{Losada2015,Losada2016,KeChen}. Nowadays, it is assumed to be the ground state for the U$_{2}$Mo compound as it proved to be more stable than the previously considered one with I4/mmm space group. Nonetheless, this can only be asserted within the limit of 3 f.u. \linebreak \newline per cell  considered in this work. At variance with U$_2$Ti \citep{Kaur}, the existence of CDW further stabilizing the $\Omega$–phase should be disregarded as no internal relaxation of coordinates was found by duplication of the cell along the c-axis, neither imaginary frequencies were observed in phonon dispersion curves \citep{Losada2016}, showing its dynamical stability.

\subsubsection{Mo-6.25 at.\%U: U$_{15}$Mo compound}
The ultimate intriguing discovery in the GS of U-Mo system is related to the new stable structure with U$_{15}$Mo composition found in this work. So far, there has been no theoretical or experimental evidence about the existence of this phase at low temperatures. 
This new structure is constructed by adding a substitutional Mo impurity to a $1\times2\times2$ supercell of $\alpha$–U \citep{Soderlind}, as depicted in Fig. \ref{fig:U15Mo}.  It is characterized by a contraction in the b-axis, compared to the $b$-parameter of $\alpha$-U and slight increase if compared to $\alpha_1$-U. Both structures lattice parameters are described in Table \ref{tb.PureUstructures}. The b-axis is the directions where the chain U-Mo-U is formed in this structure. A similar behavior is observed in the $a$-parameter. The $c$-parameter is the only one in which that behavior is reversed. These results may be deduced by observing Table \ref{tb:GSstructures}, Table \ref{tb.PureUstructures}, and Fig. \ref{fig:Displacements}. 

Another distinctive feature of U$_{15}$Mo structure with \#25 Pmm2 space group is an internal atoms movement compatible with the presence of CDWs inherited from its $\alpha_1$-U. The atomic displacements are shown in Fig. \ref{fig:Displacements}. The U$_{14}$Mo$_{2}$ compound was included in this study in order to find the driving mechanism of atomic movements, which are shown in Fig. \ref{fig:Displacements} and summarized in Table \ref{tb.distances}. It should be noticed that the presence of Mo atoms in the cell certainly affects the movement of U atoms that participate in the CDW in pure U. The displacements are lower in U$_{15}$Mo than in U$_{14}$Mo$_{2}$, being the collective movements different in both structures due to the interaction between U atoms and the nearby Mo ones. The energy gain in U$_{15}$Mo due to minimization of internal coordinates is $\Delta$E=0.1 mRy/atom. This new phase will be named $\alpha$-U$_{15}$Mo in this work.

\begin{table}[!h]
	\centering
	\begin{tabular}{|l|}
		\hline 
		\hspace{3.5cm} $\alpha$-U \\
		\hline 
		Space group: \#63 Cmcm  \\
		Lattice parameters: a=2.8181$\AA$, b=5.8426$\AA$\\
		c=4.9318 $\AA$\\
		Wyckoff possitions:  \\
		U (4c) (0,y,1/4)(0,-y,3/4)(1/2, 1/2+y,1/4)\\
		(1/2, 1/2-y,3/4)\\
		y=0.9011
		\\
		\hline 
		\hspace{3.45cm} $\alpha_1$-U \\
		\hline 
		Space group: \#62 Pnma  \\
		Lattice parameters: a=5.8426$\AA$, b=4.9318$\AA$\\
		c=5.6362 $\AA$\\
		Wyckoff possitions:  \\
		U1 (4c) (x,1/4,z)(-x,3/4,-z)\\
		(-x+1/2,3/4,z+1/2)\\
		(x+1/2,1/4,-z+1/2) x=0.3523 z=0.8667\\
		U2 (4c) x=0.8494 z=0.1165 \\
		\hline 
	\end{tabular}
	\caption{Structural information of pure U phases at low temperatures.}\label{tb.PureUstructures}
\end{table}

 
\begin{table}[]
\centering
\begin{tabular}{|c|l|}
\hline 
Phases & Distances  \\
\hline
$\alpha$-U & d=2.8181$\AA$ \\
\hline
$\alpha_1$-U & d$_1$=2.9128$\AA$, d$_2$=2.7235$\AA$\\
\hline
U$_{14}$Mo$_2$ & d$_{4-5}$=2.9398$\AA$, d$_{6-7}$=2.6763$\AA$,\\
& d$_{11-12}$=2.7248$\AA$, d$_{13-14}$=2.9882$\AA$, \\
\hline 
U$_{15}$Mo & d$_{11 -11}$=2.7735$\AA$, d$_{10-10}$=2.8322$\AA$\\
 & d$_{9-9}$=2.9465$\AA$, d$_{8-8}$=2.9225$\AA$\\
\hline
\end{tabular}
\caption{Effect of Mo addition in distances between relaxed atoms in  U$_{15}$Mo and U$_{14}$Mo, compared with $\alpha$-U and $\alpha_1$-U.}\label{tb.distances}
\end{table}
 
The internal movement in the cells of U$_{15}$Mo and U$_{14}$Mo$_{2}$ are collective and could be associated with charge density waves inherited from $\alpha$-U. This possibility and its relationship with the electronic structures of these compounds are currently under research.

A very interesting fact is the existence of structural similitudes between U$_{15}$Mo compound and the so-called $\alpha$'-phase \citep{Lehmann}. The latter has a lower limit composition of 93.75 \%at U and a contraction of b-parameter with increasing Mo content. This raises the question of whether the metastable $\alpha$'-phase is related or not with the new discovered  $\alpha$-U$_{15}$Mo one due to their composition and structural similitude, but this possibility it is not studied in this work limited to T = 0K.

\nobalance 
\section{CONCLUSIONS}\label{sec.conclusions}
This research has revealed the existence of only two compounds belonging to the U-Mo system GS: the $\Omega$- and $\alpha$-U$_{15}$Mo phases. This finding was limited to considering compounds of up to 3 times the formula unit per cell and using a supercell scheme for dilute compositions. Spin-Orbit interaction does not affect the trends and relative stability of structures in the GS of this system. 

A range of immiscibility of U in bcc-Mo based structures extended from 0 to nearby 50 at.\% of U. It is also found that the metastable GS at $x$ = 0.5 corresponds to the structure called E0 in this work. This last structure is 22 mRy/atom more stable than B2 and $\sim$ 6 mRy/atom unstable referred to the convex hull.

The $\Omega$-phase has shown no relaxations of internal atomic coordinates when duplicating the cell in the direction of the c axes, in agreement to the absence of imaginary components in the phonon spectra presented in a previous work \citep{Losada2016}.

{
The unexpectedly stable structure $\alpha$-U$_{15}$Mo, constructed by adding a substitutional Mo impurity to a $1\times2\times2$ supercell of $\alpha$-U, could be related to the metastable $\alpha$'-phase observed in quenched samples with a Mo content below 6.25 at.\% at room temperature. This is due to their common features like the composition limit and their b-parameter contraction compared to the b-parameter of $\alpha$-U. Unlike the $\Omega$-phase, U$_{15}$Mo exhibits relaxation of U positions inside the cell, resembling those occurring in $\alpha_1$-U. Similar effects were observed in metastable U$_{14}$Mo$_{2}$ compound. The presence of Mo atoms in both U$_{15}$Mo and metastable U$_{14}$Mo$_{2}$ structures affects the displacement of U atoms that participate in the CDW in pure $\alpha$-U. The spontaneous Peierls distortion observed in the last structure permeates the properties of all structures of the GS near the pure U zone of the phase. 
Furthermore, the GS of $\alpha$-U is not completely described due to the existence of phase transitions at low temperatures associated with CDWs. Consequently, it is not possible to fully determine the GS of the U$_{15}$Mo compound which is affected by those CDWs. Due to the amount of atoms involved in the CDW transitions associated to $\alpha_2$ and $\alpha_3$ phases, no attempts were made in this work to describe U$_{15}$Mo based on these structures. Therefore, in the U-rich zone, it can only be determined the structural tendencies towards the fundamental state of the U-Mo system.

The similar behavior of $\alpha$'-phase in several U-X systems (X = Nb, Ti, Zr, Ru, Re) \citep{yakel} leads to the question of whether a structure with the Pmm2 space group and composition U$_{15}$X could also be found in the GS of those systems. In addition, the existence of CDW recently found in U$_2$Ti \citep{Kaur} highlights that this could be fairly frequent phenomenon among structures with very high content of U as U$_{15}$Mo and U$_{14}$Mo$_{2}$ phases. These recent findings inspire and encourage to study the GS structures and properties of those systems.
}

\balance

{
\bibliographystyle{natbib}

\bibliography{bibliografia}

}

\end{document}